\documentclass{article} 
\usepackage{amsmath,amssymb,varioref,xcolor,paralist,graphicx,booktabs,colortbl}
\usepackage[ruled,vlined]{algorithm2e}
\usepackage{tikz}
\newcommand{\calA}{\mathcal{A}}

\newcommand{\calD}{\mathcal{D}}
\def\0{^{\phantom0}}

\newtheorem{theorem}{Theorem}

\begin{document}

\title{Association via Entropy Reduction}
\author{Anthony Gamst, Larry Wilson \and IDA Center for Communications
  Research-La Jolla \and 4320 Westerra Ct., San Diego, CA 92121 \and
  \{acgamst, larry\}@ccr-lajolla.org}

\maketitle

\begin{abstract}
Prior to recent successes using neural networks, term
frequency-inverse document frequency (tf-idf) was clearly regarded as
the best choice for identifying documents related to a query. We
provide a different score, aver, and observe, on a dataset with ground
truth marking for association, that aver does do better at finding
assciated pairs than tf-idf. This example involves finding associated
vertices in a large graph and that may be an area where neural
networks are not currently an obvious best choice.  Beyond this one
anecdote, we observe that (1) aver has a natural threshold for
declaring pairs as unassociated while tf-idf does not, (2) aver can
distinguish between pairs of documents with all terms in common based
on the frequencies of those terms while tf-idf does not, (3) aver can
be applied to larger collections of documents than pairs while tf-idf
cannot, and (4) that aver is derived from entropy under a simple
statistical model while tf-idf is a construction designed to achieve a
certain goal and hence aver may be more ``natural.'' To be fair, we
also observe that (1) writing down and computing the aver score for a
pair is more complex than for tf-idf and (2) that the fact that the
aver score is naturally scale-free makes it more complicated to
interpret aver scores.
\end{abstract}

\section{Introduction}

If we had data that told us what actors had parts in which movies, we
might hope to identify pairs of actors, maybe Bogart and Bacall or
Abbott and Costello, who worked together a lot. You might try to do
something similar with data about the co-authors of academic papers. A
natural approach would be to use the data to produce an
{\it association score} for pairs of actors with the idea that the
actors who worked together a lot would get a higher association score.

A natural first choice for the association score would be the Jaccard
index, which here would be the ratio of how many movies both actors
were in to how many movies at least one of the actors was in. This
simple score would accomplish a lot of what we want. If we were
looking to nitpick, which we will be doing here, we might say that
some of these movies are big ensemble pieces and have lots of actors
in them. It's natural to feel that appearing together in such a film
should count less towards indicating that the actors worked together a
lot. Jaccard index doesn't have any way to account for how many actors
were in each movie and so we might prefer a score with a little more
nuance.

The most commonly used association score is Term Frequency-Inverse
Document Frequency (tf-idf). Below we will say quite a lot about what
this score is, but here we just note that it is clearly accounting for
two frequencies. The Jaccard index already accounts for the frequency
of an actor appearing in movies because the denominator includes all
of those movies. The other frequency being accounted for in tf-idf is
the frequency of actors appearing in each movie, which Jaccard lacks.

We will describe tf-idf. We will then describe some properties of
tf-idf that we think aren't ideal and that therefore might encourage
one to seek another association score. We will then compare tf-idf
with our own association score, which we call aver, on a data set for
which we have some ground truth about which pairs are associated;
unfortunately, this is not data about actors in movies. After that, we
describe aver.

\section{Term Frequency-Inverse Document Frequency}

In tf-idf, the mental picture is to replace actors by documents and
movies the actors appear in by the words contained in the
documents. The words of the document matter but the order does not,
this is typically called a ``bag-of-words'' model. We want to produce
a score that measures the association of two documents by looking at
the words they have in common. Like the Jaccard index, we want the 
score to have something in the denominator that generally reduces 
the score when the document has a huge number of words. Unlike Jaccard 
index, we also want the score to be reduced when the terms are very common 
--- having ``the'' and ``and'' and ``of'' in common should not be taken as 
very much evidence of association between the two documents.

The ``term frequency'' part of tf-idf is where we put in the reduction
for the number of terms in the document. This is typically
$\mbox{tf}(t,d) = c(t,d)/N(d)$ where we are going to write $c(t,d)$
for the count of how many times the term $t$ occurs in the document
$d$ and $N(d)$ for the total number of terms in the document
$D$. Variants of the term frequency have been proposed but we stick
with this basic version. We note that while we have removed the order
of the words, we still retain the multiplicity. In our worked
comparison below, the data doesn't support multiplicities and so all
of the $c(t,d)$ will be either 0 or 1.

The ``inverse document frequency'' is where we attempt to reduce the
impact of having common terms. Here, common is going to be measured by
what fraction of the documents contain the term, that is, we ignore
multiplicities here. Typically one uses
$\mbox{idf}(t,\calD) = \log(|\calD|/M(t))$ where $\calD$ is a
collection of documents and $M(t)$ tells us how many documents contain
$t$. We instead use a standard variant, a smoothed version,
$\mbox{idf}(t,\calD) = 1 + \log(|\calD|/(M(t)+1))$. If nothing else,
this prevents us from trying to divide by 0 if we somehow had a term
$t$ that was not in any documents.

Finally, we get the full score by multiplying,
$\mbox{tf-idf}(t,d,\calD) = \mbox{tf}(t,d)\mbox{idf}(t,\calD)$.
That's a number that could be assigned to a term and a document 
in a given corpus, but it isn't an association score between
documents. In order to  compare documents, we first create a vector
associated to each document.  These vectors have components for each
term $t$. We write
$w_{d,\calD} = (\mbox{tf-idf}(t,d,\calD))_t$ to indicate that the
vector for $d$ as a part of the corpus $\calD$ has, in the $t$-th
component, the value $\mbox{tf-idf}(t,d,\calD)$. We then compute the
normalized vectors
$v_{d,\calD} = \frac{1}{||w_{d,\calD}||} w_{d,\calD}$. Finally, the
association score, which we also call tf-idf, between two documents
$d_0$ and $d_1$ of $\calD$ is given by
$\mbox{tf-idf}(d_0,d_1,\calD) = v_{d_0,\calD} \cdot v_{d_1,\calD}$.

Let us briefly observe that each $\mbox{tf-idf}(t,d,\calD)$ is
non-negative; it is 0 iff the term $t$ does not appear in the
document $d$. If we were not using smoothing, then changing the base
of the logarithm would only scale the vector $w_{d,\calD}$ and
therefore not change $v_{d,\calD}$. Because of our smoothing, the base
does matter. In our experiments below we use the natural logarithm, as
is typical.

In Table~\vref{tab:tf-idf-ex}, we compute the tf-idf association score
for the pairs of three simple documents. In this simple example, every
$\mbox{tf}(t,d)$ is either 0 if the term does not occur in the
document or else $1/4$. For terms that occur in one document, the
$\mbox{idf}(t,\calD)$ is $1.41$, in two documents is $1.00$, and in
three documents is $0.71$. This allows us to compute the various
$\mbox{tf-idf}(t,d,\calD)$ as depicted in the table.

\begin{table}
\begin{center}
\begin{tabular}{lrrrrrrr}
\toprule
$d_0$ & \multicolumn{7}{l}{a star is born} \\
$d_1$ & \multicolumn{7}{l}{the star is bright} \\
$d_2$ & \multicolumn{7}{l}{born is a verb}  \\
\midrule
& a & born & bright & is & star & the & verb \\
$w_{d_0,\calD}$ & 0.25 & 0.25 &    0 & 0.18 & 0.25 &    0 &    0 \\
$w_{d_1,\calD}$ &    0 &    0 & 0.35 & 0.18 & 0.25 & 0.35 &    0 \\
$w_{d_2,\calD}$ & 0.25 & 0.25 &    0 & 0.18 &    0 &    0 & 0.35 \\
\midrule
$v_{d_0,\calD}$ & 0.53 & 0.53 &    0 & 0.38 & 0.53 &    0 &    0 \\
$v_{d_1,\calD}$ &    0 &    0 & 0.60 & 0.43 & 0.30 & 0.60 &    0 \\
$v_{d_2,\calD}$ & 0.47 & 0.47 &    0 & 0.34 &    0 &    0 & 0.66 \\
\midrule
\multicolumn{3}{l}{$\mbox{tf-idf}(d_0,d_1,\calD)$} &
\multicolumn{4}{l}{0.34}\\
\multicolumn{3}{l}{$\mbox{tf-idf}(d_0,d_2,\calD)$} &
\multicolumn{4}{l}{0.63}\\
\multicolumn{3}{l}{$\mbox{tf-idf}(d_1,d_2,\calD)$} &
\multicolumn{4}{l}{0.10}\\
\bottomrule
\end{tabular}
\end{center}
\caption{An example of computing association via tf-idf.}
\label{tab:tf-idf-ex}
\end{table}

While there certainly are other association scores, tf-idf is quite
commonly used. For the task of finding relevant documents to user
queries, tf-idf held prime position until more recent neural network
approaches became popular. The goal of this paper is to introduce a
new association score and compare it with tf-idf on a data set for
which we have some ground truth. We therefore feel compelled to list
some properties of tf-idf that make us think that improvements might
be possible.

\begin{enumerate} \label{nitpicking tf-idf}
\item While all tf-idf association scores are between 0 and 1, they
  aren't probabilities, so it isn't so easy to interpret the
  scores. For example, there isn't a natural cut off where we would
  say that pairs that score worse than that threshold probably aren't
  associated.
\item While tf-idf is designed to reduce the impact of common words,
  if we had the documents 'a is the' and 'the is a' then they would
  get an association score of $1.0$ and be at the very top of our list
  of potentially associated pairs of documents.
\item There is no direct way to apply the tf-idf association score to
  evaluate larger collections of documents; to go back to the movie
  example, we could find neither the three stooges nor the Marx brothers
  without doing something other than applying an association score.
\item We've mentioned that there are variants of tf-idf. This is
  because it is a constructed score designed to achieve a goal and
  people can differ on how best to achieve that goal. While the work
  of Aizawa~\cite{mi} links tf-idf to the mutual information between
  the documents and the words, we might prefer something derived from
  more natural principles.
\end{enumerate}

\section{Comparing tf-idf and aver on the orkut data set}

The association score we propose, described in detail in the next
section, will address each of the issues metioned above. The goal
of this section is to whet the reader's appetite for those details
by comparing our score, aver, to tf-idf on a real-world data set
for which some ground truth is known. We hope that this anecdotal
evidence will convince the reader that aver is a valuable association
score that does do something different than tf-idf.


We chose the Orkut~\cite{orkut} dataset from the SNAP~\cite{snapnets}
dataset collection as our testbed. Orkut was a social network and
users could connect as friends. Our actors/documents will be the users
and our movies/terms will be a user and all of their friends. So now a
common term is a user with lots of friends while rarer terms are users
with fewer friends. Each term only occurs either zero or one times in
a document; either the two users are friends (or the same user) or
they are not.

The Orkut network also allowed users to create groups. The dataset
curators extracted from the 15 million plus user created groups a
collection of the ``top 5000'' groups. Our truth marking will be
whether or not the pair of users are in any one of these top 5000
groups together.

The dataset has 3,072,441 users and 117,185,083 pairs of friends. With
over 3 million users, there are over $2^{42}$ pairs of users and we
obviously can't compute that many association scores. We therefore
limited ourselves to just the pairs with over 100 friends in common;
if the two are friends, that counts as two friends in common. This
reduces us to just 5,116,585 pairs of users that need to be compared.

Of the pairs of that meet this threshold, 4,678,405 are in no top 5000
groups together while 438,180 are in a top 500 group together, only
about $8.6$ percent of these highly connected pairs of individuals are
true positives. Therefore, random selections of pairs are unlikely to be
connected and the association scores have to work pretty well to get a
better true positive rate than false positive rate.

\begin{figure}
\begin{center}
\includegraphics[width=\textwidth]{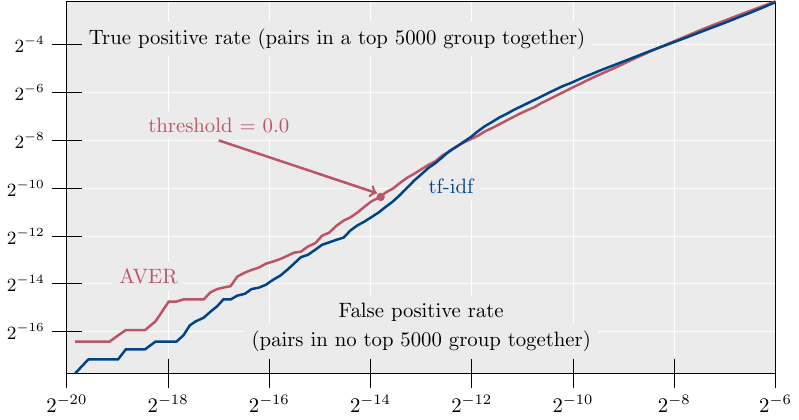}
\end{center}  
\caption{As we change the cutoff threshold for tf-idf (blue) and aver
  (red), we get different false positive and true positive rates for
  the surviving pairs. We emphasize the results for a threshold of 0
  for aver. Given a fixed false positive rate, we would prefer a
  higher true positive rate.}
\label{fig:fpr_tpr}
\end{figure}

In Figure~\vref{fig:fpr_tpr}, we plot the true and false positive
rates as we change the threshold on our two different scores, aver
(red) and tf-idf (blue). That is, given that we require tf-idf or aver
to be larger than some value $t$, we plot the fraction of the true
positives that remain and the fraction of the false positives that
remain. There are about 11 times as many false positives as true
positives, so we need the true positive rate to be something like 11
times as large for the survivors to be half true positives.

We have noted that there is no natural threshold for tf-idf and so one
needs to do something like this analysis and consider basically all
thresholds or only look at the top so many scoring pairs (as we will
do soon). For aver, there is a natural threshold, the value 0, and we
have indicated this value on the graph. With this threshold, we do
achieve the happy result that half of the remaining positives are true
positives. In general, when we allow lots of pairs (have very low
thresholds), the two scores perform more or less the same. However, we
do see a fairly noticeable difference when we get down to the lowest
false positive rates as we look at the pairs that are assigned the very
highest values for each score, where the pairs identified by aver are
more likely to be in a top 5000 group together.

Having produced an anecdote which suggests that the associations
identified by aver could be more valuable than those identified by
tf-idf, we now turn to trying to glean some insight into the
differences between the two scores. We think that the best way to do
so is to look at the pairs that score highest for each score. Pairs
that score high for both cannot tell us about differential preference,
so we will focus on the pairs that are highest scorers for one score
but not the other.

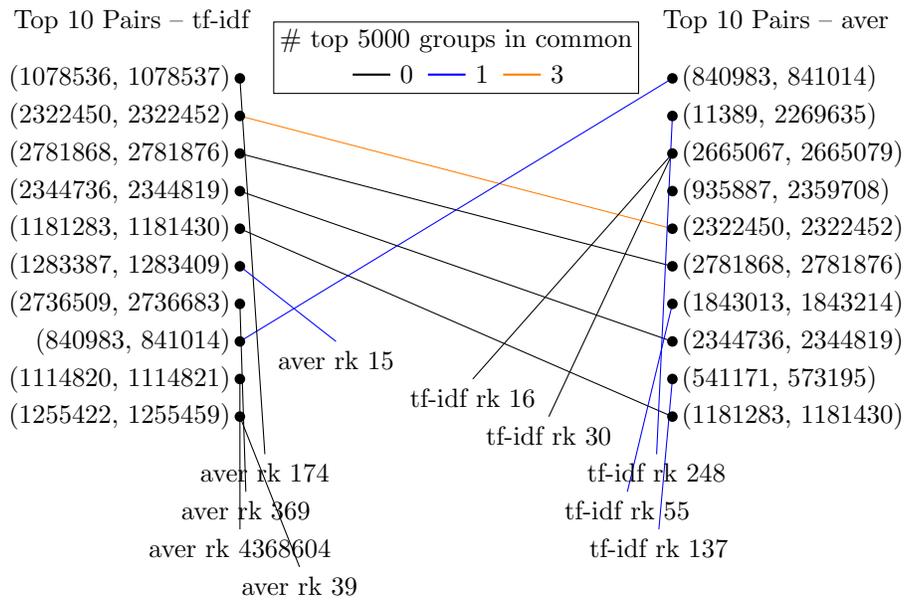
\begin{figure}
\begin{center}
\begin{tikzpicture}
\draw[orange] (0,9.5) -- (5.75,8);
\draw (0,9) -- (5.75,7.5);
\draw (0,8.5) -- (5.75,6.5);
\draw (0,8) -- (5.75,5.5);
\draw[blue] (0,6.5) -- (5.75,10);

\draw (0,5.5) -= (0.793, 3.5) node [below] {aver rk 39};
\draw (0,6) -- (0.001, 4) node [below] {aver rk 4368604};
\draw[blue] (5.75,6) -- (5.570, 4) node [below, black] {tf-idf rk 137};
\draw[blue] (5.75,7) -- (5.151, 4.5) node [below, black] {tf-idf rk 55};
\draw (0,7) -- (0.079, 4.5) node [below] {aver rk 369};
\draw (0,10) -- (0.332,5) node [below] {aver rk 174};
\draw[blue] (5.75,9.5) -- (5.540,5) node [below, black] {tf-idf rk 248};
\draw (5.75,9) -- (4.107, 5.5) node [below] {tf-idf rk 30};
\draw (5.75,9) -- (3.096, 6) node [below] {tf-idf rk 16};
\draw[blue] (0,7.5) -- (1.278,6.5) node [below, black] {aver rk 15};

\fill (0,10) node [left] {(1078536, 1078537)} circle (2pt);
\fill (0,9.5) node [left] {(2322450, 2322452)} circle (2pt);
\fill (0,9) node [left] {(2781868, 2781876)} circle (2pt);
\fill (0,8.5) node [left] {(2344736, 2344819)} circle (2pt);
\fill (0,8) node [left] {(1181283, 1181430)} circle (2pt);
\fill (0,7.5) node [left] {(1283387, 1283409)} circle (2pt);
\fill (0,7) node [left] {(2736509, 2736683)} circle (2pt);
\fill (0,6.5) node [left] {(840983, 841014)} circle (2pt);
\fill (0,6) node [left] {(1114820, 1114821)} circle (2pt);
\fill (0,5.5) node [left] {(1255422, 1255459)} circle (2pt);
\fill (5.75,10) node [right] {(840983, 841014)} circle (2pt);
\fill (5.75,9.5) node [right] {(11389, 2269635)} circle (2pt);
\fill (5.75,9) node [right] {(2665067, 2665079)} circle (2pt);
\fill (5.75,8.5) node [right] {(935887, 2359708)} circle (2pt);
\fill (5.75,8) node [right] {(2322450, 2322452)} circle (2pt);
\fill (5.75,7.5) node [right] {(2781868, 2781876)} circle (2pt);
\fill (5.75,7) node [right] {(1843013, 1843214)} circle (2pt);
\fill (5.75,6.5) node [right] {(2344736, 2344819)} circle (2pt);
\fill (5.75,6) node [right] {(541171, 573195)} circle (2pt);
\fill (5.75,5.5) node [right] {(1181283, 1181430)} circle (2pt);

\draw (0.25,10.75) node[left] {Top 10 Pairs -- tf-idf};
\draw (5.5,10.75) node[right] {Top 10 Pairs -- aver};

\draw (2.875,10.5) node {\# top 5000 groups in common};
\draw[thick] (1.5,10.05) -- (2,10.05) node [right] {0};
\draw[thick,blue] (2.5,10.05) -- (3,10.05) node [right, black] {1};
\draw[thick,orange] (3.5,10.05) -- (4,10.05) node [right, black] {3};
\draw (0.45,9.8) rectangle (5.3,10.75);

\end{tikzpicture}
\end{center}
\caption{The 10 highest scoring pairs in tf-idf (left) and aver
  (aver). The five pairs that occur in both top 10s are connected, the
  other five pairs (for each) point to where they would be in the
  other list if it went far enough. The color of the line reflects the
  number of top 5000 groups containing both of the users in the pair.}
\label{fig:top_10}
\end{figure}

In Figure~\vref{fig:top_10}, we provide the user ids for the top 10
scoring pairs under each association score. It turns out that five of
these pairs are in common; so we will ignore them. Some of the
remaining score fairly highly in the other score (tf-idf's 6th highest
score is the 15th highest in aver) while others were not so close to
the top 10 (tf-idf's 9th highest score is aver's 4368604th
highest). The figure also includes information about truth marking;
three of the five pairs that were top 10 in aver but not tf-idf were
in a top 5000 group together while only one of the five pairs that
were top 10 in tf-idf but not aver.

We now begin to look more closely at the pairs that were in one of the
top 10 lists but not the other. First, both of the top two scores in
tf-idf were 1.0 because, for both of those pairs, both individuals had
exactly the same set of friends. It is therefore very interesting that
aver ranks these two pairs as 5 and 174 indicating that aver finds
them very different. In this case, that worked out very well because
the pair that aver ranked 5 is in 3 of the top 5000 groups together
while the other pair isn't in any. While this is merely an anecdote,
it does suggest that tf-idf could have some blind spots and that not
all pairs of documents that have exactly the same set of terms should
be treated exactly the same way.

\begin{table}
\begin{center}
\begin{tabular}{l|rrrrr}
\toprule
tf-idf rank & 1 & 6 & 7 & 9 & 10 \\
aver rank & 174 & 15 & 369 & 4368604 & 39 \\
\# top 5000 groups & 0 & 1 & 0 & 0 & 0 \\
degree(first) & 180 & 140 & 204 & 753 & 217 \\
degree(second) & 180 & 146 & 203 & 776 & 212 \\
\# common nbrs & 180 & 138 & 194 & 723 & 202 \\
Avg deg common nbrs & 947.7 & 148.1 & 190.7 & 699.0 & 219.5 \\
Med deg common nbrs & 217.0 & 131.0 & 183.5 & 500.0 & 107.0 \\
\midrule
tf-idf rank & 248 & 30 & 16 & 55 & 137 \\
aver rank & 2 & 3 & 4 & 7 & 9 \\
\# top 5000 groups & 1 & 0 & 0 & 1 & 1 \\
degree(first) & 415 & 123 & 128 & 109 & 127 \\
degree(second) & 327 & 125 & 127 & 120 & 131 \\
\# common nbrs & 311 & 114 & 119 & 103 & 112 \\
Avg deg common nbrs & 58.3 & 115.6 & 103.9 & 90.3 & 95.4 \\
Med deg common nbrs & 42.0 & 71.5 & 75.0 & 69.0 & 60.5 \\
\bottomrule
\end{tabular}
\end{center}
\caption{Details about the five pairs that were in the top ten scores
  for tf-idf but not AVER and vice-versa.}
\label{tab:only one}
\end{table}

We provide some details about the five pairs that had top ten
scores in tf-idf but not aver in the top half of
Table~\vref{tab:only one}. We begin with the rankings under the two
different scores and then the number of top 5000 groups containing
both users; this is also available in Figure~\ref{fig:top_10}. Then, we
provide the number of friends of each of the two users in the order
given in the first table, which is numerical order; recall that we
consider everyone to be friends with themselves. We think that the two
scores use the global frequency of these terms differently, so we have
provided some summary statistics of the frequencies of the terms, the
average frequency and the median frequency. The bottom half of the
table has the same information for the five pairs that had top ten
scores in aver but not tf-idf.

For the pair that got a score of 1.0 in tf-idf but which aver ranked
much lower, it is clear that the friends that they have in common have
large numbers of friends and so this is the sort of thing that tf-idf
wants to downplay but cannot. Of the five that only made the top ten
for tf-idf, the only one that actually was in a top 5000 group
together was the one with the smallest number of common neighbors and
those neighbors with the smallest average degree (global
frequency). Similarly, for those that made the top ten of aver but not
tf-idf, it was the three with the lowest average degree for common
neighbors that were in top 5000 groups together. It seems that aver
does a better job of getting to pairs with less common terms in common.

Our reading of this is that the aver score is more sensitive to the
term frequency than tf-idf is while being more tolerant of having
terms not in common. Presumably sometimes that will be beneficial and
sometimes it will not, depending on the intended application.

\subsection{Applying aver to groups larger than pairs}

We have noted that tf-idf only applies to pairs of
documents and thus can't find the three stooges or the Marx
brothers. 
aver can be applied to tuples of arbitrary length.

There were already too many pairs to score all of them, we
certainly don't want to try to score all triples of users. We required
our starting pairs to have 100 friends in common; for larger sets we
will require at least 10 friends common to the whole set. Given a set
(starting with one of the one million highest scoring pairs), we found
every user we could add to the set and still meet the threshold on
number of common friends and computed the aver score for this larger
set. If any of the larger sets exceeded the score of the starting set,
we take the larger set with the greatest score and carry on from
there; if not, we record the starting set and its score.

Of the resulting one million sets, the highest aver score came from a
set of 29 users, their user ids can be found in
Table~\vref{tab:top29}. While none of the top 5000 groups contains all
of these users, several contain large subsets of them. The table gives
the sizes of the top 5000 groups that contain the highest numbers of
this set of users. These are fairly small groups, under 100 users,
containing at least 15 of our set of 29. We take that as pretty good
evidence that this group is strongly associated. None of these 29
users were involved in any of the top ten pairs of scores for aver, so
it really was the common friendships of the large group that makes
them stand out. We invite the reader to imagine how they would have
tried to use tf-idf to find a similarly highly-connected set of users.

\begin{table}
\begin{center}
\begin{tabular}{rrrrrr}
\toprule
1174444 & 1284130 & 1315039 & 1322325 & 1629325 & 1673907 \\
1674318 & 1674540 & 1674721 & 1674841 & 1674884 & 1675009 \\
1675157 & 1675174 & 1675378 & 1675439 & 1675529 & 1675581 \\
1675885 & 1675891 & 1675917 & 1675997 & 1676002 & 1676113 \\
1676118 & 1676207 & 1676245 & 1760238 & 2469768 \\
\end{tabular}
\begin{tabular}{lrrrrr}
\midrule
Group size & 95 & 80 & 54 & 84 & 99 \\
\# of these in group & 26 & 18 & 16 & 16 & 15 \\
\bottomrule
\end{tabular}
\end{center}
\caption{The 29 users who make up the top scoring set of Orkut users that we
found by the algorithm described in the text. Also, the group size and
number of these 29 in the five top 5000 groups containing the most of these
29 users.}
\label{tab:top29}
\end{table}

\section{Our association score -- aver}

We propose a score, called aver, based on entropy reduction.
In order to describe the score, we have to discuss the probability
model on which it is based, but briefly:
We compute the entropy of the observed data. We then treat
seriously the claim that the pair (or larger set) of documents are
associated by treating their common terms as coming from a new
document of their collaboration and their remaining terms (if any) as
being due to their own individual efforts. We compute the entropy of
this altered data set and our score is the reduction in entropy of
the modified document set (which could be negative; that is, the
entropy could increase). For large data 
sets, we wouldn't want to compute the second entropy from scratch, 
so we provide a (not particularly pretty) formula that computes the 
score. Because the score was not constructed, we have to prove that
(in the general case that the collaboration doesn't account for a 
large proportion of the terms) the aver score for a set of 
documents decreases (that is the indication of association is less) as
the number of times a common term is seen increases. We end with some
reasonable criticisms of aver.

\subsection{The rank-one model and entropy}

We have some documents that consist of some terms. Let's write
$c(t,d)$ for the count of how often term $t$ occurs in document
$d$. We write $N := \sum_{t,d} c(t,d)$. We also will need
$T(t) := \sum_d c(t,d)$ which counts how often term $t$ occurs across
the entire corpus and $D(d) := \sum_t c(t,d)$ which tells us how many
terms occur in document $D$.

We imagine that these data arise from a very simple statistical
model. We have a probability distribution over terms, $p(t)$, and a
probability distribution over documents, $q(d)$. Our model is that the
data was generated by choosing a term from $p$ and a document from $q$
and then adding term $t$ to document $d$. In the bag-of-words
interpretation of documents, this suffices. The MLE is $p(t) = T(t)/N$
and $q(d) = D(d)/N$. The advantage of a simple statistical model is
that we impose very little of our own preconceptions on the data.

The entropy of the data is
$$E = -\left(\sum_t p(t)\log(p(t))+\sum_d q(d)\log(q(d))\right)$$
The entropy would be maximized if all of the terms had the same
probability and all of the documents had the same
probability. We assume that our data has some structure, that there
are subjects which the documents cover and infrequent words which are
confined to those subjects. Modeling these features would result in
a distribution with lower entropy. We hope to identify associated 
actors by observing that their actions in common correspond to useful 
structure and so lead to a further reduction in entropy.

\subsection{Collaboration}

We want to find that Abbott and Costello appeared in a bunch of movies
together and also the three stooges and the Marx brothers. We think
that, really, the individual actors weren't hired, but rather the team
was hired. We're going to say that the common movies were due to the
collaboration of the actors rather than to the individual actors.

Rewriting that in the language of documents containing terms, we want
to claim that the common terms of a set $\calA$ of documents were due
to collaboration rather than being the work of those individual
documents. Therefore, we introduce a new document, $d_\calA$ that
captures the work resulting from the collaboration. We remove those 
common terms from the individual documents that belong to
$\calA$. Previously, we had $c(t,d)$ to represent the count of the
number of times the term $t$ occurs in document $d$. Let's use $c'$
for the counts in this imaginary world in which the collaboration
is accounted for in this way.

Then, $c'(t,d_\calA)$ is supposed to represent the common use of the
term $t$ among the documents in $\calA$ and so should be
$c'(t,d_\calA) = \min_{d\in\calA} c(t,d)$. We remove the work due to
the collaboration from the individual documents in $\calA$. Therefore,
if $d\in\calA$, for every term $t$, we have
$c'(t,d) = c(t,d)-c'(t,d_\calA)$. For all documents outside of $\calA$
and for all terms, $c'(t,d) = c(t,d)$. From these new counts we
compute the MLE rank-one model and compute its entropy, let's call that
$E'$. Then, our aver score for the documents $\calA$ is
$$\mbox{aver}(\calA,\calD) = E-E',$$ 
the reduction of entropy when we associate
the common actions to the collaboration.

To use the formula above to compute $E'$, we really need to look at
all of the documents in the whole corpus and all of the terms. It
would certainly be preferable to have a formula that only touched the
documents in $\calA$ and the terms common to them, those with
$c'(t,d_\calA) > 0$. We will provide such a formula and leave it to
the reader to verify that it gives the same value. Let's first discuss
some additional notation.

We had $N = \sum_{t,d} c(t,d)$ so naturally $N' = \sum_{t,d} c'(t,d)$
and we'd like to emphasize that these values are different because
we've reduced the number of times that the terms common to $\calA$
appear (provided $\calA$ has more than one document). We want $D'(d)$
to be the number of terms in document $d$. We have
$D'(d_\calA) = \sum_t c'(t,d_\calA)$ and then for documents in
$\calA$ we have $D'(d) = D(d) - D(d_\calA)$. For terms $t$, the number
of appearances from each $d \in \calA$ is reduced by $c'(t,d_\calA)$
but they do appear that many times in the new document $d_\calA$, so
we have $T'(t) = T(t)-(|\calA|-1)c'(t,d_\calA)$. In particular, we can
find that $N' = N - (|\calA|-1)D'(d_\calA)$.

Finally, we note that
$$E = 2\log(N) - \frac{1}{N}\left(\sum_t T(t)\log(T(t)) + \sum_d D(d)\log(D(d))\right)$$
and we will write $e = -N(E-2\log(N))$ and assume that we have computed
this value. Then,
\begin{eqnarray*}
  \mbox{aver}(\calA,\calD) &=& 2\log(N/N') + \frac{N-N'}{NN'} e\\
  && \quad - \frac{1}{N'} \sum_{t:c'(t,d_\calA) \ne 0}
  T(t)\log(T(t))-T'(t)\log(T'(t)) \\
  && \quad - \frac{1}{N'} \sum_{d\in\calA}
  (D(d)\log(D(d))-D'(d)\log(D'(d))\\
  && \quad + \frac{1}{N'} D'(d_\calA)\log(D'(d_\calA))
\end{eqnarray*}

\subsection{Computing aver on our simple example}

We would like to compute aver over the very simple corpus of
Table~\vref{tab:tf-idf-ex}; see Table~\vref{tab:aver-ex}. This forces
us to confront the fact that we haven't specified the base of the
logarithm in these entropy counts. For that corpus, we're going to
choose to use natural logarithms. Obviously any other choice of base
just scales the scores (and doesn't change the signs provided we choose
a base greater than 1). In the table, we directly compute $E$ and
$E'$. In this discussion, we will compute $e$ and $e'$ and from them
$E$ and $E'$.

We have $D(d_i) = 4$ for each $i$ and so
$\sum_d D(d)\log(D(d)) = 3\cdot4\cdot\log(4) \approx 16.64$. There
are three terms that occur once (and we note that $\log(1)=0$), three
that occur twice, and one that occurs three times, so
$\sum_t T(t)\log(T(t)) = 3\cdot2\log(2)+3\cdot\log(3) \approx 7.45$.
Therefore, $e \approx 24.09$ and hence $E = 2\log(12) - e/12 \approx 2.96$.

When we compute the aver score for $\calA = \{d_0,d_1\}$, we have that
the collaboration $d_\calA$ produces ``star is''  and leaves the two
documents with two terms. Also, ``star'' now only occurs once and
``is'' only occurs twice. This makes $N'=10$. We have
$\sum_d D'(d)\log(D'(d))=3\cdot2\log(2) + 4\cdot\log(4) \approx 9.70$.
Remembering again that $\log(1)=0$, we have that
$\sum_t T'(t)\log(T'(t))=3\cdot2\log(2) \approx 4.16$. This gives
that $e' \approx 13.86$ and $E' = 2\log(10) - e'/10 \approx 3.22$ and
therefore that $\mbox{aver}(\{d_0,d_1\},\calD) \approx -0.26$.

In Table~\vref{tab:aver-ex}, we see that aver gives a different
ordering to the three pairs. We also see that aver gives all of the
pairs a negative score, which indicates that aver does not find it
likely that the pairs are associated. This might be due to the small
size of the corpus, the small size of each document, or the limited
global variability.  We also see that we have
computed a score for the triple of all three documents, which we could
not do directly using tf-idf. In this case, where the three documents
only have the very common word ``is'' in common, aver gives this
larger collaboration an even worse score than any of the pairs. It is
not at all unusual to have negative aver scores when positing a
collaboration between documents that are not associated.
We see in Figure~\vref{fig:fpr_tpr} that setting a
threshold of $0.0$ for the aver score cuts out quite a lot of pairs in
that data and that data only includes pairs of users that have at
least 100 friends in common.

\begin{table}
\begin{center}
\begin{tabular}{lrrclrr} \toprule
&&& \, & $d'_0$ & \multicolumn{2}{l}{a born} \\
\raisebox{1.5ex}[0pt]{$d_0$} &
\multicolumn{2}{l}{\raisebox{1.5ex}[0pt]{a star is born}} &&
$d'_1$ & \multicolumn{2}{l}{the bright} \\
\raisebox{1.5ex}[0pt]{$d_1$} &
\multicolumn{2}{l}{\raisebox{1.5ex}[0pt]{the star is bright}} &&
$d'_\calA$ & \multicolumn{2}{l}{star is} \\
\raisebox{1.5ex}[0pt]{$d_2$} &
\multicolumn{2}{l}{\raisebox{1.5ex}[0pt]{born is a verb}} &&
$d'_2$ & \multicolumn{2}{l}{born is a verb} \\
\midrule
&&& & $d'_0$ & 2/10 & -0.32 \\
\raisebox{1.5ex}[0pt]{$d_0$} & \raisebox{1.5ex}[0pt]{4/12} &
\raisebox{1.5ex}[0pt]{-0.37} && $d'_1$ & 2/10 & -0.32 \\
\raisebox{1.5ex}[0pt]{$d_1$} & \raisebox{1.5ex}[0pt]{4/12} &
\raisebox{1.5ex}[0pt]{-0.37} && $d'_\calA$ & 2/10 & -0.32 \\
\raisebox{1.5ex}[0pt]{$d_2$} & \raisebox{1.5ex}[0pt]{4/12} &
\raisebox{1.5ex}[0pt]{-0.37} && $d'_2$ & 4/10 & -0.37 \\
\raisebox{1.5ex}[0pt]{a} & \raisebox{1.5ex}[0pt]{2/12} &
\raisebox{1.5ex}[0pt]{-0.30} && a & 2/10 & -0.32 \\
\raisebox{1.5ex}[0pt]{born} & \raisebox{1.5ex}[0pt]{2/12} &
\raisebox{1.5ex}[0pt]{-0.30} && born & 2/10 & -0.32 \\
\raisebox{1.5ex}[0pt]{bright} & \raisebox{1.5ex}[0pt]{1/12} &
\raisebox{1.5ex}[0pt]{-0.21} && bright & 1/10 & -0.23 \\
\raisebox{1.5ex}[0pt]{is} & \raisebox{1.5ex}[0pt]{3/12} &
\raisebox{1.5ex}[0pt]{-0.35} && is & 2/10 & -0.32 \\
\raisebox{1.5ex}[0pt]{star} & \raisebox{1.5ex}[0pt]{2/12} &
\raisebox{1.5ex}[0pt]{-0.30} && star & 1/10 & -0.23 \\
\raisebox{1.5ex}[0pt]{the} & \raisebox{1.5ex}[0pt]{1/12} &
\raisebox{1.5ex}[0pt]{-0.21} && the & 1/10 & -0.23 \\
\raisebox{1.5ex}[0pt]{verb} & \raisebox{1.5ex}[0pt]{1/12} &
\raisebox{1.5ex}[0pt]{-0.21} && verb & 1/10 & -0.23 \\
\midrule
$E$ && 2.96 && $E'$ && 3.22 \\
\midrule
\multicolumn{2}{l}{$\mbox{aver}(\{d_0,d_1\},\calD)$} & $-0.26$ \\
\multicolumn{2}{l}{$\mbox{aver}(\{d_0,d_2\},\calD)$} & $-0.14$ \\
\multicolumn{2}{l}{$\mbox{aver}(\{d_1,d_2\},\calD)$} & $-0.23$ \\
\multicolumn{2}{l}{$\mbox{aver}(\{d_0,d_1,d_2\},\calD)$} & $-0.29$ \\
\bottomrule
\end{tabular}
\end{center}
\caption{Computing the entropy for the original small data set and for
the case where we propose a collaboration among the documents in
$\calA = \{d_0,d_1\}$ using natural logarithms. We also give the aver
scores for all three pairs of documents and for the triple of documents.}
\label{tab:aver-ex}
\end{table}



\subsection{How aver depends on the frequency of terms}

tf-idf is designed so that having a more frequent term as a common
term contributes less to the score. aver operates differently: A
simple statistical model is assumed, the entropy of the observed
data under the fitted model is computed, and the association 
between a set of actors is scored in terms of the change in 
entropy that assuming that association entails.
Although it was not specifically designed to capture such
effects, it would still be nice to know that the aver score 
is stronger when less frequent terms are in common.

In particular, we'd like to know that
$\frac{\partial \mbox{aver}(\calA,\calD)}{\partial T(t)} < 0$ for a
term $t$. In a complicated data situation like this, other values do
depend on $T(t)$ but we're going to simplify things and treat all the
variables that appear in the aver formula as independent. Let us also
simplify things by using natural logarithms; any other base just
scales the scores and so doesn't change the sign of any derivatives.

Now
$T'(t) = T(t)-(|\calA|-1)c'(t,d_\calA)$; we write
$m = (|\calA|-1)c'(t,d_\calA)$ where $m$ is a constant so that
$T'(t) = T(t)-m$. With these choices, $\mbox{aver}(\calA,\calD)$ is a
constant plus
$$\frac{-1}{N}T(t)\log(T(t)) + \frac{1}{N'}(T(t)-m)\log(T(t)-m)$$
which includes the contribution of $T(t)$ to $e$ if we are thinking in
terms of the formula; this is obvious when we think of the difference
of entropies.

For $f(x) = x\log(x)$, the derivative $f'(x) = \log(x)+1$. Therefore,
$$\frac{\partial \mbox{aver}(\calA,\calD)}{\partial T(t)} = \frac{-1}{N}(\log(T(t))+1) + \frac{1}{N'}(\log(T(t)-m)+1)$$
Scaling by $NN'$, we observe that we need to find the sign of
$$-N'(\log(T(t))+1) + N(\log(T(t)-m)+1).$$
We can find the value of $N'$ that makes this be 0. As
the coefficient of $N'$ is negative, the derivative is negative for
larger values of $N'$. The derivative is 0 when
$$N' = N\frac{\log(T(t)-m)+1}{\log(T(t))+1}.$$
We will think of $N'$ as $N - s$ where $s$ is related to the number of
terms in the collaboration. The derivative is negative when
$$N-s > N(\log(T(t)-m)+1)/(\log(T(t))+1)$$ and hence when
$$-s > N(-1+(\log(T(t)-m)+1)/(\log(T(t))+1)$$ or
$$s < N\frac{\log(T(t))-\log(T(t)-m)}{\log(T(t))+1}.$$

\begin{theorem}
Provided that the collaboration accounts for a sufficiently small
fraction of the total observed data, that is, if
$$(|\calA|-1)D'(d_\calA) < N\frac{\log(T(t))-\log(T(t)-m)}{\log(T(t))+1}$$
then $$\frac{\partial \mbox{aver}(\calA,\calD)}{\partial T(t)} < 0$$
when we hold all other terms in the formula constant.
\end{theorem}

\subsection{Concerns about tf-idf and aver}

We earlier enumerated four concerns about tf-idf that made us think it
might be worth looking for another association score. Let's start with
some concerns about aver.

\begin{enumerate}
\item The formula for aver is complex and therefore difficult to
  comprehend; the score is almost a black-box. Also, it might be
  difficult to code efficiently.
\item We can scale the scores arbitrarily by changing the base of the
  logarithm and therefore the actual returned values from aver can't
  really be sensibly interpreted.
\end{enumerate}

On the other hand, we do feel that we have at least made some answer
to the four comments we made earlier
(page~\pageref{nitpicking tf-idf}) about tf-idf.


\begin{enumerate}
\item Association is supposed to be recognized by a reduction in
  entropy so there is a natural cutoff of 0 for whether pairs are
  associated; naturally we expect some non-causally connected pairs to
  get positive scores, but perhaps these will be clustered around
  0. For a score that can be scaled arbitrarily, 0 is the only
  possible natural threshold.
\item In the orkut data, we found aver distinguishing two cases where
  the documents had all of their terms in common, giving one a top-ten
  score and the other a much lower score.
\item We have seen that we can apply the aver score to collections
  larger than pairs. However, because it requires actions in common by
  the whole collection, it won't find communities where most of the
  actors do the common actions. We do think, however, it will identify
  closely connected subsets of these communities and, perhaps, from
  that base, one can find the whole community.
\item While tf-idf is a constructed score, we claim that aver, being
  derived from entropy under a simple statistical model, is a more
  natural thing; in particular, it should be less susceptible to a
  proliferation of variations. There was human involvement in choosing
  how to change the data to account for collaboration, so presumably
  other humans could make other choices, but we believe that the basic
  idea of having a reduction in entropy measure association is solid.
\end{enumerate}

\bibliography{aver}
\bibliographystyle{plain}

\end{document}